\documentstyle[preprint,aps,psfig]{revtex}

\draft
\tightenlines
\begin{document}


\title{ 
Transitions between metastable states in silica clusters
	}
\author{ 
       Claudio Brangian
	}
\address{ 
	  Institut f\"ur Physik, Johannes Gutenberg Universit\"at,
          Staudinger Weg 7, D-55099 Mainz, Germany
	  }
\author{Oreste Pilla and Gabriele Viliani}
\address{ 
         Dipartimento di Fisica and Istituto Nazionale di Fisica della
         Materia, Universit\`a di Trento, I-38050 Povo, Trento, Italy
        }
\maketitle

\begin{abstract}
\noindent
Relaxation phenomena in glasses can be related to jump processes between
different minima  of the potential energy in the configuration space.
These transitions play a key role in the low temperature regime, giving rise 
to tunneling systems responsible for the anomalous specific heat and thermal 
conductivity in disordered solids with respect to crystals. 
By using a recently developed numerical algorithm, we study the potential
energy landscape of silica clusters, taking as a starting point the location
of first order saddle points. This allows us to find a great number of 
adjacent minima. We analyze the degree of cooperativity  
of these transitions and the connection of physical
properties with the topology of the configuration space.
We also identify two-level systems (pairs of minima constituting a 
tunneling system) and calculate the quantum mechanical ground state
splitting by means of the WKB approximation.    
\end{abstract}


\begin{section}{Introduction}
The theoretical investigation of the properties of disordered systems is a
very difficult task, because  the lack of simmetry usually prevents
analytical approaches.
Many authors have found it convenient to rely on the concept of 
{\sl potential energy landscape} in the configuration space 
(see e.g. Frauenfelder {\it et al.} 1991, Berry 1993, Heuer and Silbey 1993,
Mohanty {\it et al.} 1994, Heuer 1997, Angelani {\it et al.} 1998, 
Mousseau and Barkema 1998, Wales {\it et al} 1998).

From a purely qualitative point of view, we may imagine the 
multidimensional analogue of a surface rich in climbing points (first order
saddles) and valleys (the various minima). A detailed analysis of the 
energy landscape topology requires numerical simulation.
\par
In this work we analyze the potential energy landscape of SiO$_2$ clusters, 
namely [SiO$_2$]$_{20}$, [SiO$_2$]$_{30}$, and [SiO$_2$]$_{50}$, which are
 comparable in size to
(or larger than) monoatomic and binary systems so far investigated. 
After a short description of the numerical methods used to move up and down
the multidimensional hypersurface, we report our results regarding:
\begin{enumerate}
\item  the structure of the clusters (resembling that of the bulk solid,
 although affected by surface effects);
\item the energy distribution of the stationary points of the potential 
 energy function;
\item a physical interpretation of geometrical quantities uselful
     to characterize transitions between different minima;
\item the identification of tunneling systems.
\end{enumerate} 
\end{section}

\begin{section}{Numerical methods}
We simulate the interaction among Si and O atoms
with the recently developed pair potential by Van Beest {\it et al.} (1990), 
modified with a very short range contribution (Guissani and Guillot 1996),
 necessary to avoid
unphysical divergencies. It is given by
\begin{equation}
\Phi_{ij}={q_i q_j\over r_{ij}}+ a_{ij}\exp{(-b_{ij}r_{ij})}-
{c_{ij}\over r_{ij}^6}+4\epsilon_{ij}\bigg[
\bigg({\sigma_{ij}\over r_{ij}}\bigg)^{24}-
\bigg({\sigma_{ij}\over r_{ij}}\bigg)^6 \bigg]\;,
\label{BKSpot}
\end{equation} 
where both  {\sl i} and {\sl j} indexes run on all Si and O atoms, and the
values of the parameters were determined by Van Beest {\it et al.} (1990) and
Guissani and Guillot (1996).
This potential has already been widely used (with slight 
modifications) in molecular dynamics studies of the liquid and glassy
phases of
silica (see e.g. Vollmair {\it et al.} 1996, Horbach {\it et al.}
1996, Taraskin and Elliott 1997).
As our purpose is to simulate clusters, we used free boundary conditions.
We adopted the procedure described in detail in Daldoss
 {\it et al.} (1998, 1999) to locate
minimum-saddle-minimum triplets, that are the key ingredients
 to describe jump
processes. Schematically the steps are:
\begin{itemize}
\item descent towards a minimum by the conjugate gradient method 
 (Press {\it et al.} 1986), starting from a randomly chosen configuration; 
\item ascent towards the vicinity of a saddle, following the
eigenvector corresponding to the lowest eigenvalue;
\item once the
potential energy along the path of the previous item starts decreasing, we
take the corresponding configuration ({\it hill-climbing point}, Berry 1993)
as the starting point for a Newton-Raphson search (Press {\it et al.} 1986)
of the first order saddle point; 
\item a descent on both sides of the saddle following again the
 eigenvector of the minimum eigenvalue.
\end{itemize}			
This technique provides approximate {\it
adjacent} minima that are subsequently fed into the Newton algorithm for
accurate location. With this numerical procedure we found many thousand
minimum-saddle-minimum triplets (hereafter
 {\sl double well potentials}, DWP).
\end{section}

\begin{section}{Results}  

\begin{subsection}{The Structure of the clusters}
We obtain information on the structure of the cluster by the radial 
distribution function $g(r)$, defined in the same way as for extended
systems: it shows the bond lengths, the short range order, and  also
allows for the calculation of the co-ordination number. The results are
reported in Fig.1, which was obtained by averaging
over the various configurations obtained (both minima and saddles)
to have good statistics. We have plotted the partial $g(r)$ for the
different
bonds (Si-Si, Si-O, O-O) and cluster sizes. The resulting bond
lengths obtained by the main peaks are in good agreement with
the experimental data on bulk structures (see table \ref{legami})
and with simulations
with periodic boundary conditions (Taraskin and Elliott 1997).
Nevertheless we observe smaller peaks on the low-distance side of the
nearest-neighbour peak in the Si-Si and O-O bonds. In our opinion
this is due to surface effects: in fact it appears from the
figure that these anomalies tend to disappear with increasing system size.
Very satisfactory is also the result concerning the co-ordination number:
it is 4 for Si-O bond, in agreement with the tetrahedral structure
typical of SiO$_2$. This is also confirmed by the analysis of
Si-Si, also equal to 4, and O-O, equal to 6.
\end{subsection}
\begin{subsection}{Energy distributions of stationary points}
We now analyze the topological features of the potential energy
hypersurface; at first we present the energy distribution of the
stationary points (saddles and minima) that form double wells.
In Fig. 2 we report the three energy distributions for saddles, lower
minima and upper minima of the various DWP, respectively; only the result
for 150 atoms is presented, the situation being quite analogous in the other 
two cases. 
We note that the distributions are very similar,
with defined peaks superimposed onto a broad backgroud; the peaks are due to 
the fact that certain configurations are favoured for given values of the 
binding energy, and so they act like attraction basins during the
descent towards the minima. The presence of these structures in the
distribution is
more evident with increasing system size: in [SiO$_2$]$_{20}$ the curve looks 
smoother (Brangian 1998). It should be remarked that Fig. 2
does not refer to the {\it total} distribution of stationary points,
because their number increases
exponentially with the system size: we sampled partially the configuration
space, and we are quite confident not to have introduced
systematic errors in
this sampling. The only thing we can note is that maybe our numerical 
investigation is in some way prevented from reaching very low lying (that is 
crystalline-like) configurations: since we are not interested in a careful
thermodynamical analysis (Doye and Wales 1998), but only in transitions
between stable disordered states, we think this fact does not
constitute a serious drawback.
\end{subsection}

\begin{subsection}{Topological features of double well potentials}
Double well potentials can be characterized by many quantities: the first
one is
the {\it asymmetry} $\Delta$, that is the energy difference between the 
two connected minima of a DWP. This parameter is essential for the
identification of candidate {\it two level systems} (TLS), which
require a value of $\Delta$ of the order of less than $\approx$1 K (however,
as we will see, this condition is not sufficient to
identify a TLS). Our results indicate that the asymmetries
are distributed exponentially, the most part being lower than 5000 K;
the distribution
is not sensitive to the system size. Equally important are the {\it
energy barriers}
$V$, i.e. the energy differences between a minimum and the corresponding
saddle. Of course for every  DWP there are two values of $V$, one for
the relaxation process and one for the activation. The distribution
of the relaxation barriers similar in shape to that of the $\Delta$'s,
but on average the values of $V$ are smaller, being significantly present
only up to 1000 K.
There seems to be little correlation between the asymmetry and
the barrier (both for  the activation and relaxation).

We have evaluated also the (mass weighted) {\it euclidean distance}
between pairs of minima, defined as
\begin{equation}
dist(a,b)=\bigg[\sum_i\frac{m_i}{\tilde{m}}
|{\bf r}_{i,a}-{\bf r}_{i,b}|^2\bigg]^{1/2}\;,
\label{de}
\end{equation}
where $a$ and $b$ are the two minima, $m_i$ the
 single atom mass, ${\bf r}_{i,\alpha}$ its position
 in the $\alpha$ configuration, and ${\tilde{m}}=\sum_im_i/N$.
In Fig. 3 we report the relative distributions.
The distributions do not extend
very much beyond $\approx$10$\sigma$ and present a maximum
at $\approx 2\sigma$.  

Another very interesting parameter to consider is
the {\it participation number} defined as
\begin{equation}
N_{\rm part}=\sum_i \frac{d_i^2}{d_{\rm max}^2}.
\label{numpart}
\end{equation}
Here $d_i$ refers to the atom-atom distances in (\ref{de})
and $d_{max}$ is the distance of the atom that moves most during
the transition; the partecipation number gives an idea on the numbers of
atoms involved in a transition. In Fig. 4(a) we report this quantity for
the three cases studied; the same quantity normalized to the number of
particles constituting the clusters is reported in Fig. 4(b). We see
that $ N_{\rm part}$ is a nearly scaling quantity with $N$, indicating
that at least an appreciable part of the atoms that move in the
transitions belong to the bulk.

We can make use of {\it multidimensional transition state theory} (for a
review see Hanggi 1985) to estabilsh a link between the potential
energy landscape and the (classical) relaxation dynamics of the system.
Under appropriate conditions (Hanggi 1985) the classical, thermally
activated transition probability between two metastable minima is given by
\begin{eqnarray}
\Gamma=\nu^{\ast}\exp{\bigg({-E_b\over \kappa_B T}\bigg)} \\
\nu^{\ast}={\prod_{i=1}^N \nu_i^M \over \prod_{j=1}^{N-1} \nu_j^S} \;,
\end{eqnarray}
where $\nu^{\ast}$ is an effective frequency that takes into account the
effect
of all the degrees of freedom on the one which forces the transition.
This frequency can be rewritten as
\begin{equation}
\nu^{\ast}={\nu_0\over R}
\end{equation}
with $\nu_0$ the lowest frequency of the dynamical matrix in the starting
minimum, while $R$ is the product of all the $N-1$ positive eigenvalues of
the dynamical matrix at the saddle point,
divided by the corresponding product at the minimum. In Fig. 5 we show
the value of this {\it entropic factor} $R$ (which enhances or reduces the
transition
probability), versus the energy barrier value, both for the relaxation and
the activation
transitions. This plot evidences a rather marked correlation between $R$ and
the barrier height, and in most cases $R < $1 or even $\ll $1.
\end{subsection}

\begin{subsection}{Two-level systems (TLS)}
We describe now the most difficult part of our work, that is the selection
of {\it two-level systems}. Following the energy landscape paradigm, these
are DWP that imply purely quantum mechanical relaxation processes. The
calculation has been carried out by assuming the validity of the 1D
Wentzel-Kramers-Brillouin (WKB) approximation (Froman and Froman 1965,
Landau and Lifchitz 1967; Schiff 1968; a discussion on the validity of this
approximation in the case of Ar clusters, as well as on the effects to be
expected when it is released, can be found in Daldoss {\it et al.} 1999):
the splitting of the ground state is given
in terms of the action integral between the two minima. In principle, in
order to apply the WKB procedure it would be necessary to find the least
action path, i.e. the classical path that takes from one minimum to the
other and minimizes the action integral; this involves rather heavy numerical
calculations. As a starting point, we decided to use the path that takes
from one minimum to the other, and that is defined at each point by the
direction of the minimum eigenvector, as described in Section II. Work on
the minimization
of the action integral is in progress (Brangian {\it et al.} 1999).
It should be stressed that as cumbersome as this procedure may look, it
is probably the simplest way to get a quantitative estimate of the
tunneling splitting. It should also be mentioned that, in agreement with
previous works (Heuer and Silbey 1993, Heuer 1997, Daldoss {\it et al.}
1998), we find roughly one TLS for 1000 DWP's, which implies a very
extensive search strategy.

We have performed an {\it a posteriori} test on the reliability of the
use of the 1D WKB scheme. The main hypothesis is that, along the chosen
1D path (the least-action or a very close one), the degrees of freedom other
than the considered one are decoupled from it: in this case it is
reasonable to expect that the Schroedinger equation may be (nearly) factorized
into 3$N$ mutually independent equations, a condition for the applicability
of the WKB scheme in many dimensions (Schiff 1968). If this
factorization should actually take place, the eigenvalues of the
dynamical matrix (which are proportional to the curvatures of the
hypersurface) other than the lowest one should remain constant along the
chosen path. The departure of these eigenvalues from constancy gives a
measure (though a qualitative one) of the invalidity of the 1D scheme.
In Fig. 6 we show, as an example, the results for two TLS belonging to
clusters of 90 atoms. In one case (Fig. 6(b)) the approximation is
satisfied in a
very good way; on the contrary, in the other case (Fig. 6(a)) there
is appreciable mixing of the low
energy eigenvectors. These results are in qualitative agreement with those
relative to Ar clusters (Daldoss {\it et al}, 1999); they imply that the actual
structure of TLS in disordered systems is probably much more complex than
expected, since many-dimensional effects seem to play important roles.

\end{subsection}  
\end{section}
\begin{section}{Conclusions}
We have reported preliminary results of an extensive investigation on the
properties of the potential energy landscape in SiO$_2$ clusters of
three different sizes (60,90, and 150 atoms); the aim of this research is
to find connections between the properties of the landscape and the high-
and low-temperature relaxation dynamics. By analyzing the structure of the
clusters and the topological features of their energy landscape (and
in particular of its minima and first-order saddle points), we have
identified tunneling centres and   studied the conditions of validity of
the WKB approximation, which allows a quantitative estimate of the
tunneling splitting.
\end{section}

\begin{center}ACKNOWLEDGMENTS\\
This work was supported in part by the Parallel Computing Initiative of
the INFM.
\end{center}

\begin{center}
REFERENCES\\
\end{center}
\noindent
Angelani, L., Parisi, G., Ruocco, G., and Viliani, G., 1998, {\it
Phys. Rev. Lett.}, {\bf 81,} 4648.

\noindent   
Berry, R.S., 1993, {\it Chem. Rev.}, {\bf 93,} 2379.

\noindent
Brangian, C., 1998, {\it Thesis} (University of Trento).

\noindent
Brangian, C., Pilla, O., and Viliani, G., 1999, to be published.

\noindent
Daldoss, G., Pilla, O., and Viliani, G., 1998, {\it Phil. Mag.} B,
{\bf 77,} 689.

\noindent
Daldoss, G., Pilla, O., Viliani, G., Brangian, C., and
Ruocco, G., 1999, to appear on
{\it Phys. Rev.} B.

\noindent
Doye, J.P.K., and Wales D.J., 1998 {\it Phys. Rev. Lett.}, {\bf 80,} 1357.

\noindent
Frauenfelder, H., Sligar, S.G., and Wolynes, P.G.,
1991, {\it Science}, {\bf 254,} 1594.

\noindent
Froman, N., and Froman, O.O., 1965, {\it JWKB Approximation} (North-Holland:
Amsterdam).

\noindent   
Guissani, Y., and Guillott, B., 1996, {\it J. Chem. Phys.}, {\bf 104,} 7633.

\noindent
Hanggi, P., 1986, {\it J. Stat. Phys.}, {\bf 42,} 105.

\noindent 
Heuer, A., 1997, {\it Phys. Rev. Lett.}, {\bf 78,} 4051.

\noindent
Heuer, A., and Silbey, R.J., 1993, {\it Phys. Rev. Lett.}, {\bf 70,} 3911.

\noindent
Horbach, J., Kob, W., Binder, K., and Angell, C.A., 1996, {\it
Phys. Rev.} E, {\bf 54,} R5827.

\noindent
Landau, L., and Lifchitz, E., 1967, {\it M\'ecanique Quantique}, chapter
7 (MIR: Moscow).

\noindent  
Mohanty, U., Oppenheim, I., and Taubes, C.H., 1994, {\it
Science}, {\bf 266,} 425.

\noindent
Mousseau, N., and Barkema, G.T., 1998, {\it Phys. Rev.} E, {\bf 57,} 2419.                    

\noindent
Press, W.H., Flannery, B.P., Teukolsky, S.A., and  Vetterling, W.T., 1986,
{\it Numerical Recipes} (Cambridge University Press: Cambridge).

\noindent
Schiff, L.J., 1968, {\it Quantum mechanics} (McGraw-Hill: New York).

\noindent   
Taraskin, S.N., and Elliott, S.R., 1997, {\it Phys. Rev.} B, {\bf
56,} 8605.

\noindent   
Van Beest, B.W.H., Kramer, G.J., and  Van Santen, R.A., 1990, {\it Phys. Rev.
Lett.}, {\bf 64,} 1955.

\noindent   
Vollmair, K., Kob, W., and Binder, K., 1996 {\it Phys. Rev.}
B {\bf 54,} 15808.

\noindent
Wales, D.J., Miller, M.A., and Walsh, T.R., 1998, {\it Nature},
{\bf 394,} 758.

\vskip 1cm
\noindent
\narrowtext
\begin{table}
\caption{{\sl Comparison between experimental bond lenght
obtained in vitreous
silica and those calculated from our clusters.}}
\centering
$\begin{array}{|ccc|} \hline\hline
\qquad\qquad & d_{\rm exp}\; [\AA ]& d_{\rm num}\; [\AA] \\\hline\hline
Si-O & 1.61 & 1.62\\\hline
O-O & 2.63 & 2.61 \\\hline
Si-Si &3.13 & 3.04  \\
\hline\hline
\end{array}$
\label{legami} 
\end{table}

FIGURE CAPTIONS\\
\vskip 1cm
Fig. 1. Radial distribution function $g(r)$ in clusters of
three different sizes. Besides the three main peaks, which are also
found in bulk systems, we notice
other peaks for the O-O and Si-Si bonds, probably due to the presence
of surface co-ordination defects.
\vskip 1cm
Fig. 2. Energy distribution for saddles, upper and lower minima
 in clusters of 150 atoms.
\vskip 1cm
Fig. 3. Distribution of euclidean distances between connected minima,
in units of $\AA$
and $\sigma=1.6 \AA$, i.e. the Si-O bond length.
\vskip 1cm
Fig. 4. Participation number: the $y$ axis units are chosen such that the
integrals of
the various curves is equal to 1. The bottom plot refers to the
participation ratio normalized to the number of atoms.
\vskip 1cm
Fig. 5. Entropic ratio $R$ as a function of the barrier height for
$N=$60, 90, and 150.
\vskip 1cm
Fig. 6. Variation of the 10 lowest eigenvalues of the dynamical matrix 
along the minimum eigenvalue path (see text) in two TLS of clusters of
90 atoms. The minima are arbitrarily assigned the $\pm$1
values of the coordinate along the path.

\newpage
\begin{figure}[h]
\psfig{file=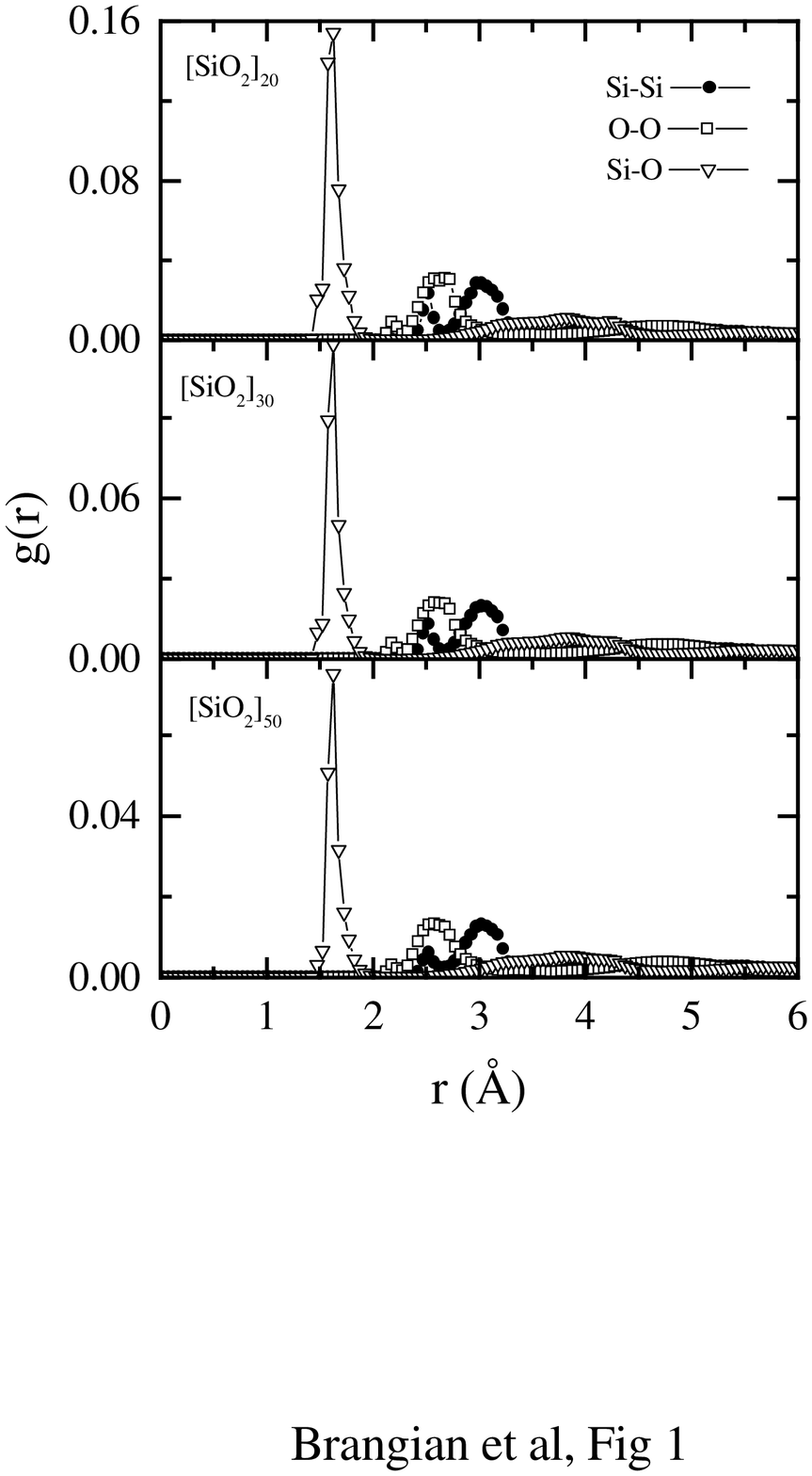,width=13cm,height=22cm}
\end{figure}

\newpage
\begin{figure}[h]
\psfig{file=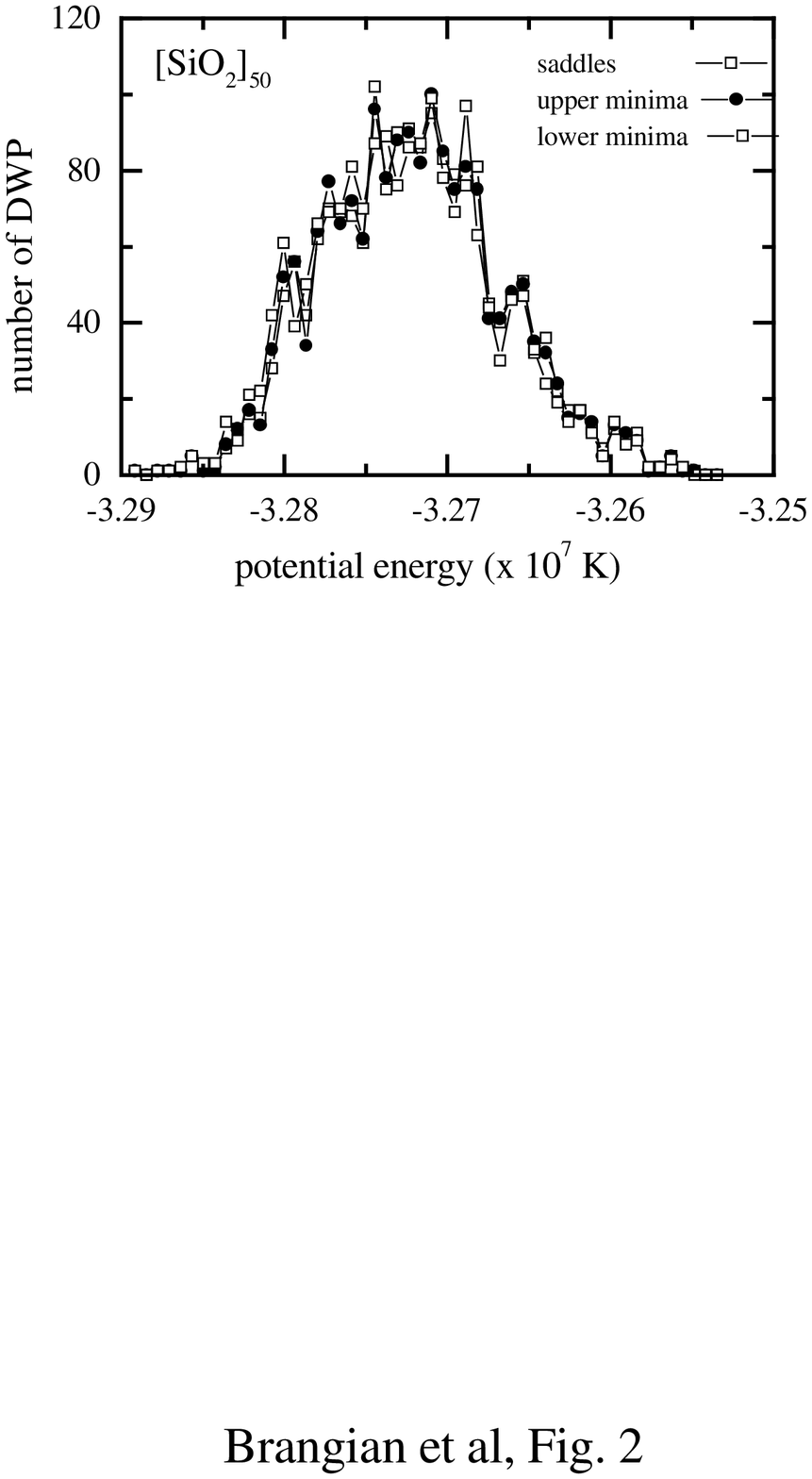,width=13cm,height=22cm}
\end{figure}

\newpage
\begin{figure}[h]
\psfig{file=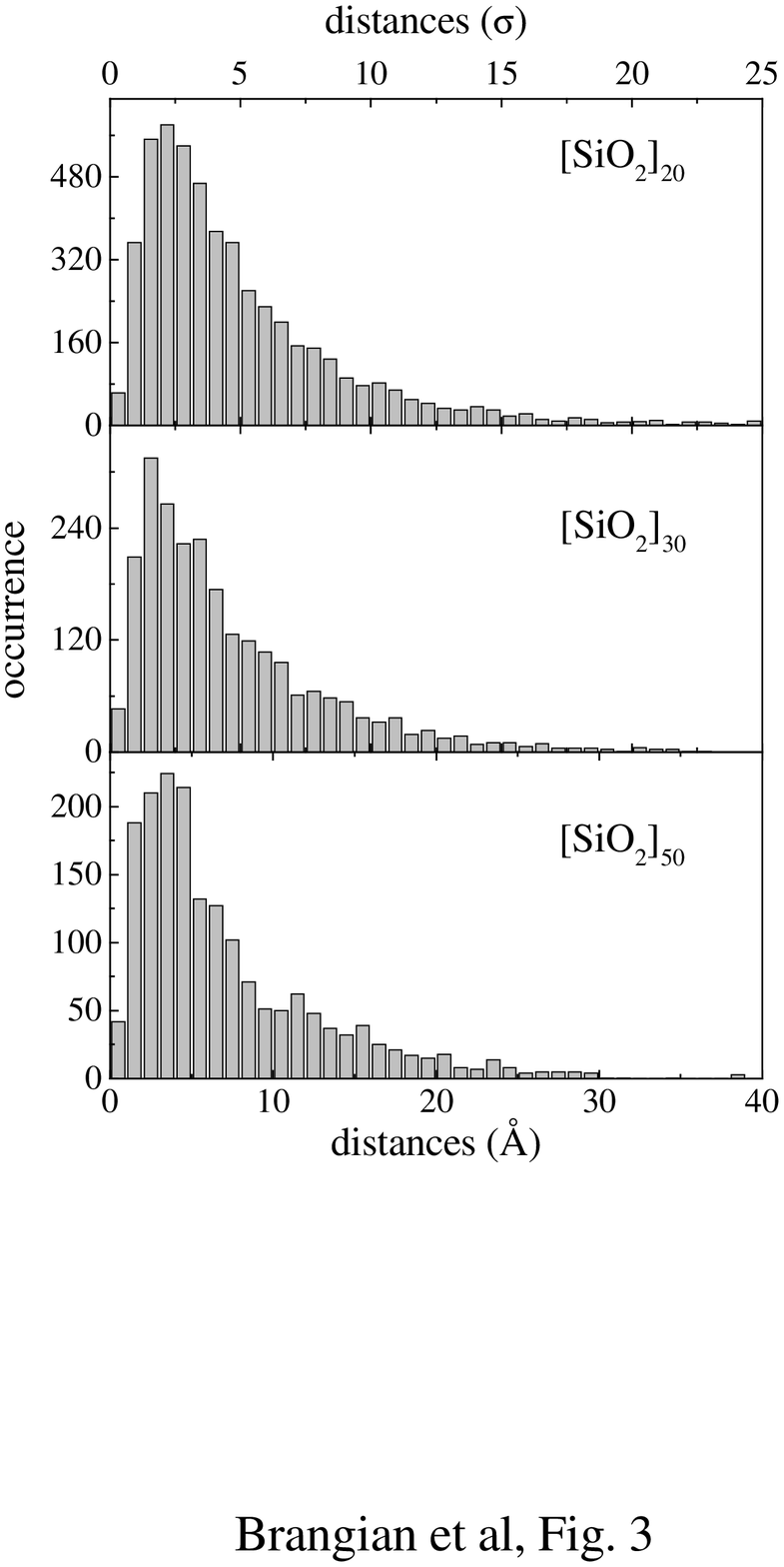,width=13cm,height=22cm}
\end{figure}

\newpage
\begin{figure}[h]
\psfig{file=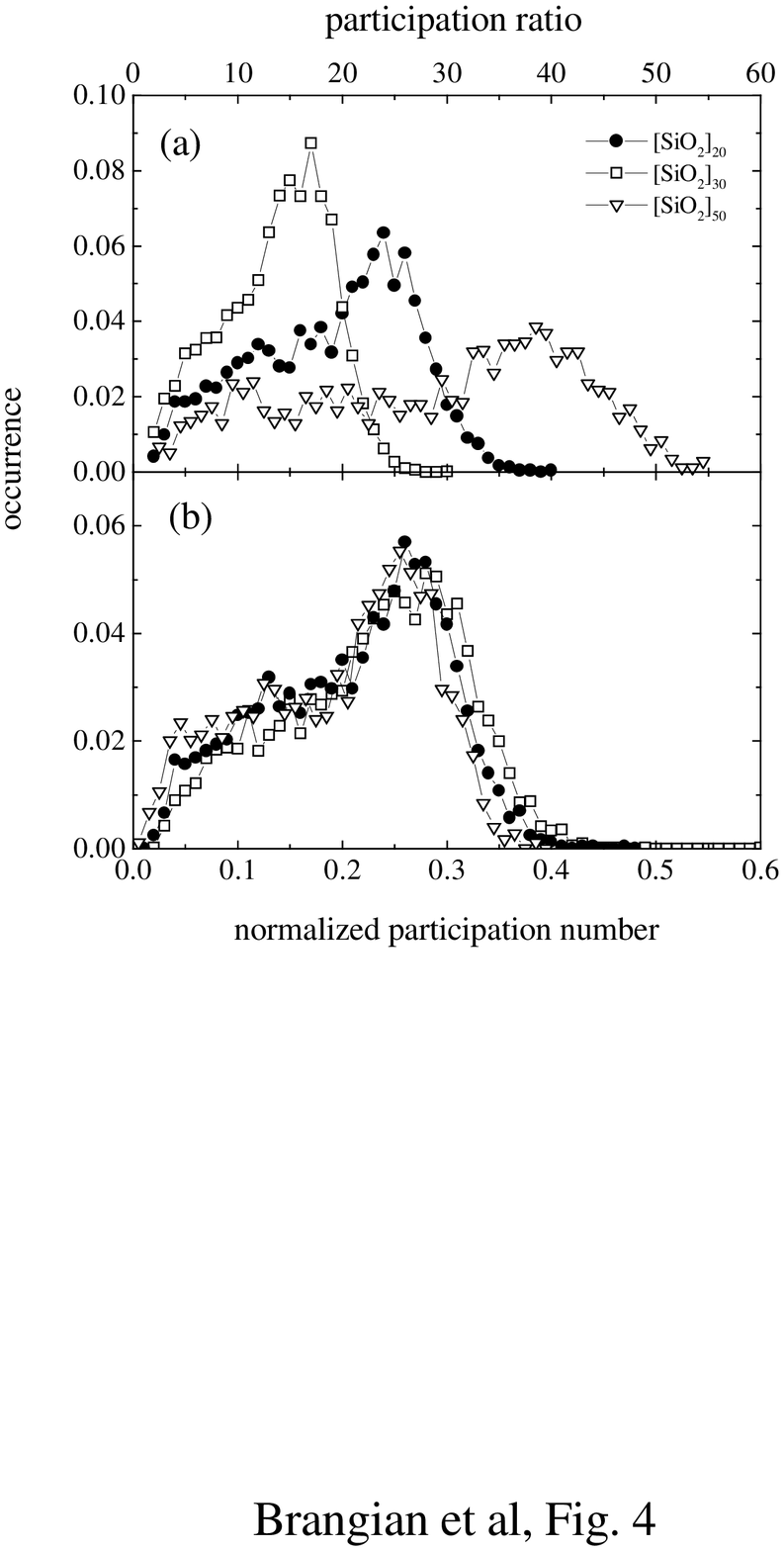,width=13cm,height=22cm}
\end{figure}

\newpage
\begin{figure}[h]
\psfig{file=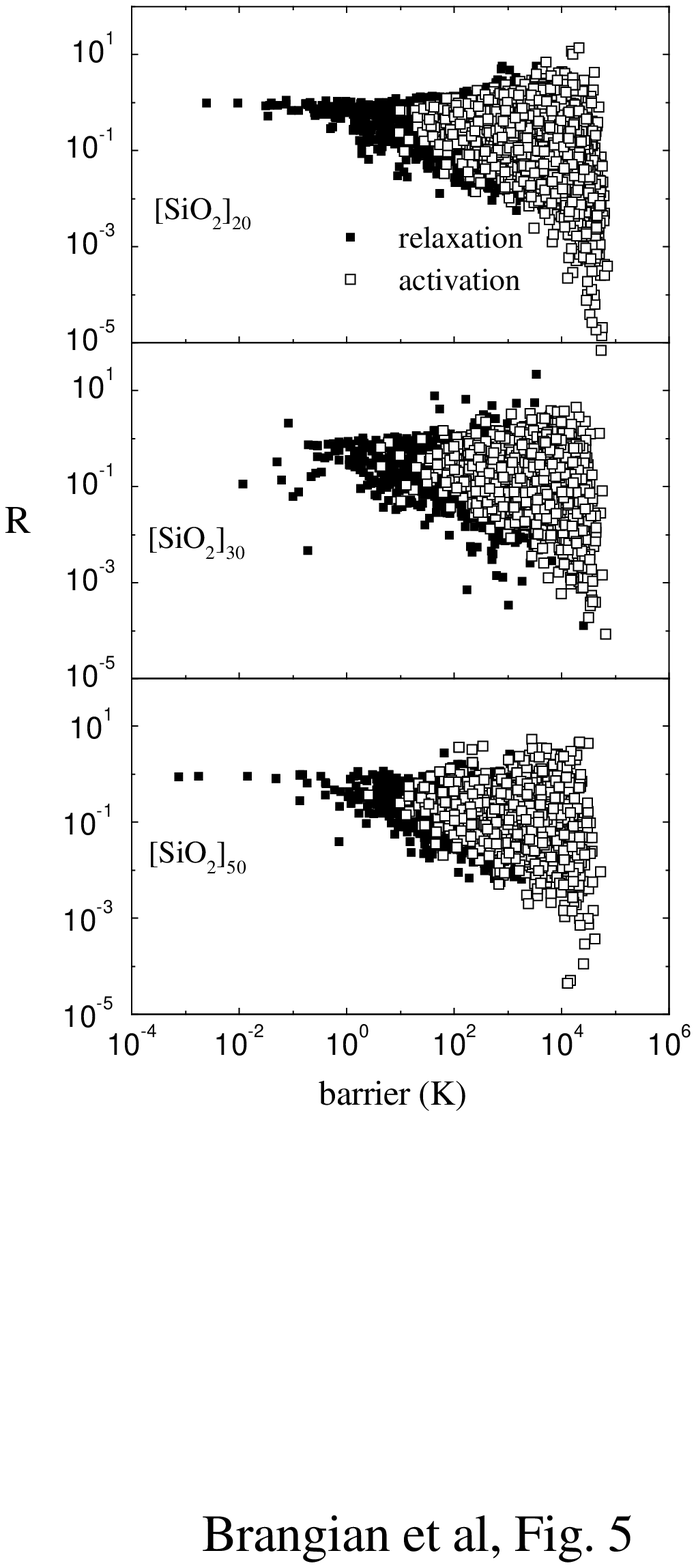,width=13cm,height=22cm}
\end{figure}

\newpage
\begin{figure}[h]
\psfig{file=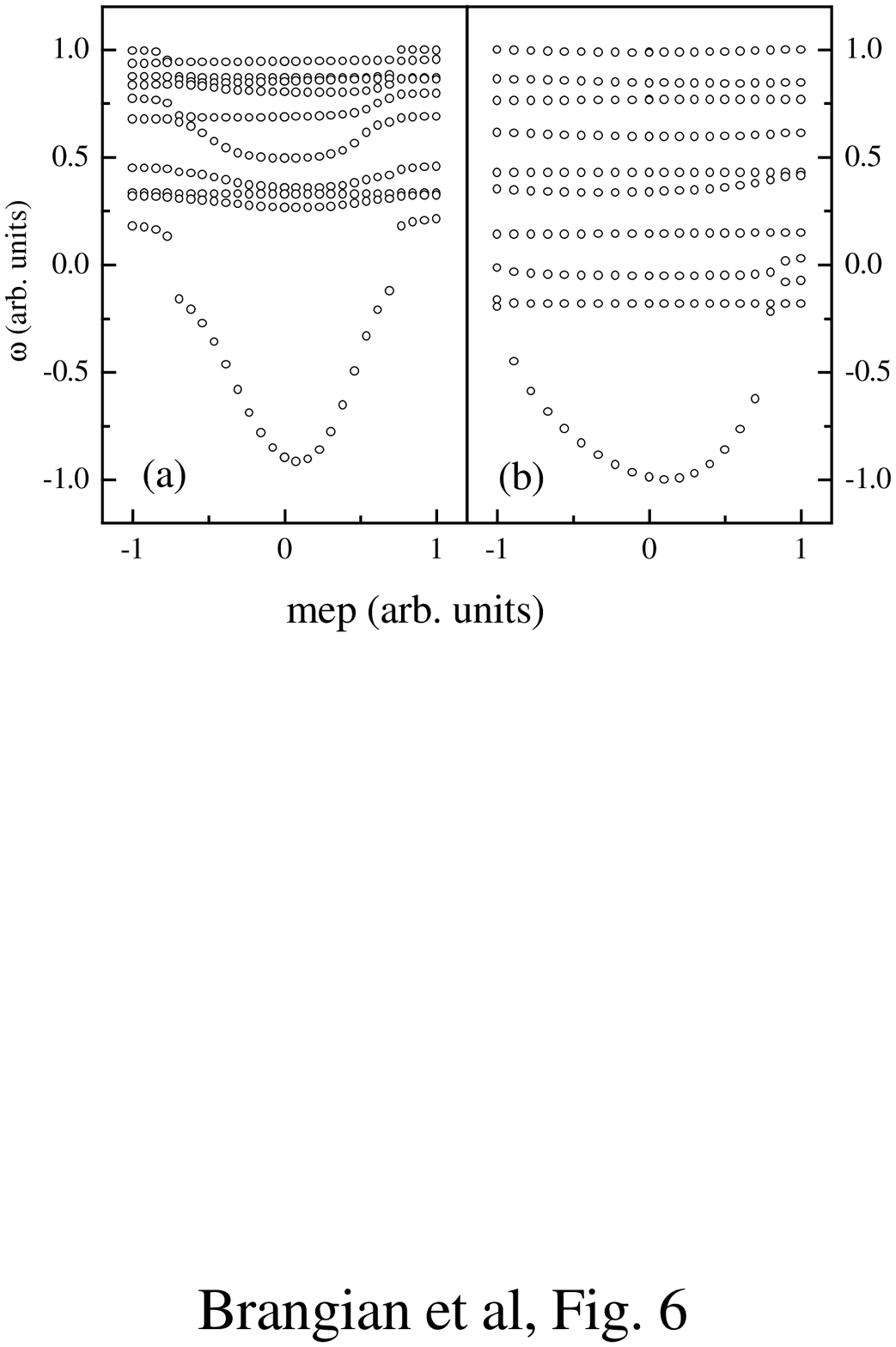,width=13cm,height=22cm}
\end{figure}
                                                                      
\end{document}